\documentclass[journal=jacsat,manuscript=article]{achemso}

\usepackage{amsmath}
\usepackage{graphicx}
\usepackage{float}
\usepackage{xcolor}
\usepackage[normalem]{ ulem }
\usepackage{soul}

\author{A. De Cecco}
\affiliation{Univ. Grenoble Alpes, CNRS, Grenoble INP, Institut N\' eel, Grenoble, France}
\author{V.\,S. Prudkovskiy}
\affiliation{Univ. Grenoble Alpes, CNRS, Grenoble INP, Institut N\' eel, Grenoble, France}
\alsoaffiliation{School of Physics, Georgia Institute of Technology, Atlanta, USA}
\author{D. Wander}
\author{R. Ganguly}
\affiliation{Univ. Grenoble Alpes, CNRS, Grenoble INP, Institut N\' eel, Grenoble, France}
\author{C. Berger}
\affiliation{Univ. Grenoble Alpes, CNRS, Grenoble INP, Institut N\' eel, Grenoble, France}
\alsoaffiliation{School of Physics, Georgia Institute of Technology, Atlanta, USA}
\author{W.\,A.\,de\,Heer}
\affiliation{School of Physics, Georgia Institute of Technology, Atlanta, USA}
\alsoaffiliation{TICNN, Tianjin University, China}
\author{H. Courtois}
\author{C.\,B. Winkelmann}
\affiliation{Univ. Grenoble Alpes, CNRS, Grenoble INP, Institut N\' eel, Grenoble, France}
\email{clemens.winkelmann@neel.cnrs.fr}

\title{Non-invasive nanoscale potentiometry and ballistic transport in epigraphene nanoribbons}

\begin{document}

\begin{abstract}

The recent observation of non-classical electron transport regimes in two-dimensional materials has called for new high-resolution non-invasive techniques to locally probe electronic properties. We introduce a novel hybrid scanning  probe technique to map the local resistance and electrochemical potential  with  nm- and $\mu$V resolution, and we apply it to study epigraphene nanoribbons grown on the sidewalls of SiC substrate steps. Remarkably, the potential drop is non uniform along the ribbons, and $\mu$m-long segments show no potential variation with distance. The potential maps are in excellent agreement with measurements of the local resistance. This reveals ballistic transport in ambient condition, compatible with micrometer-long room-temperature electronic mean free paths.

\end{abstract}

Recent experiments have demonstrated that epigraphene nanoribbons grown on the sidewalls of SiC substrate steps \cite{Sprinkle2010,Ruan2012} present one-dimensional ballistic transport properties involving a single quantum channel of conductance with electronic mean free paths exceeding 10 $\mu$m even at room temperature \cite{Baringhaus2014}. These confirmed properties \cite{Aprojanz2018,Baringhaus2015} are in contrast with the Coulomb blockade or disordered-induced insulating behavior observed in etched ribbons produced from exfoliated graphene \cite{Han2010,Gallagher2010,Droscher2011}. They have no equivalent in any other material than pristine carbon nanotubes \cite{Frank1998}, and are still not well understood. Transport scenarios involve the topologically protected graphene electronic edge state and multibody  interaction in  charge-neutral graphene \cite{Baringhaus2014,Li2016}.  An important question is the  sensitivity of the properties to the graphene quality and ribbon edge disorder as  ballistic transport was observed so far only for epigraphene sidewall ribbons, produced directly in shape at high temperature ($\sim$800 C) and subsequently either annealed in ultra-high vacuum (UHV) or protected with  alumina.
Other unconventional electronic transport mechanisms have also been reported in 2D graphene, such as hydrodynamic flow of the electron fluid even up to relatively high temperatures \cite{Crossno2016,Bandurin2016}, outlining the rich physics of the system and the need to further investigate these new non-classical electron transport regimes. Such studies should optimally combine local high-resolution electronic measurements and structural studies.
 
In this Letter, we present high-resolution local resistance measurements as well as 2D maps of the local potential of sidewall epigraphene nanoribbons. Using a scanning probe approach combining atomic force microscopy (AFM) and scanning tunneling microscopy (STM), we both map the potential landscape, as the nanoribbon is voltage-biased along its main axis, and perform local resistance measurements. Most notably, and in contrast with usual STM-based potentiometry mesurements, the technique can be applied to a mixed conducting / insulating surface. The potential (resistance) profile  is quasi-independent of distance along extended ribbon segments, indicating larger than $\mu$m-long mean free paths in these quasi-neutral ribbons, even with no high-temperature UHV annealing.

The graphene nanoribbons were epitaxially grown on inclined nanofacets resulting from the annealing of trenches etched in a 4H-SiC substrate \cite{Sprinkle2010,Ruan2012, Baringhaus2014} (see Supp. Info). This technique produces ribbons $\sim$ 20 -- 100 nm wide with well defined edge termination \cite{Ruan2012,Baringhaus2015,Palacio2015,Celis2016}. The samples consist of a large number (about a hundred) of mm-long epigraphene nanoribbons  connected in parallel. Two extended Ti/Au contacts, about 30 $\mu$m apart and perpendicular to the nanoribbons, were evaporated through a SiN stencil mask to avoid  organic resists and limit surface contamination. After contact deposition, high-temperature annealing is no longer possible as it would lead to diffusion of the gold contact on the substrate.

\begin{figure}[t]
\includegraphics[width=0.8\columnwidth]{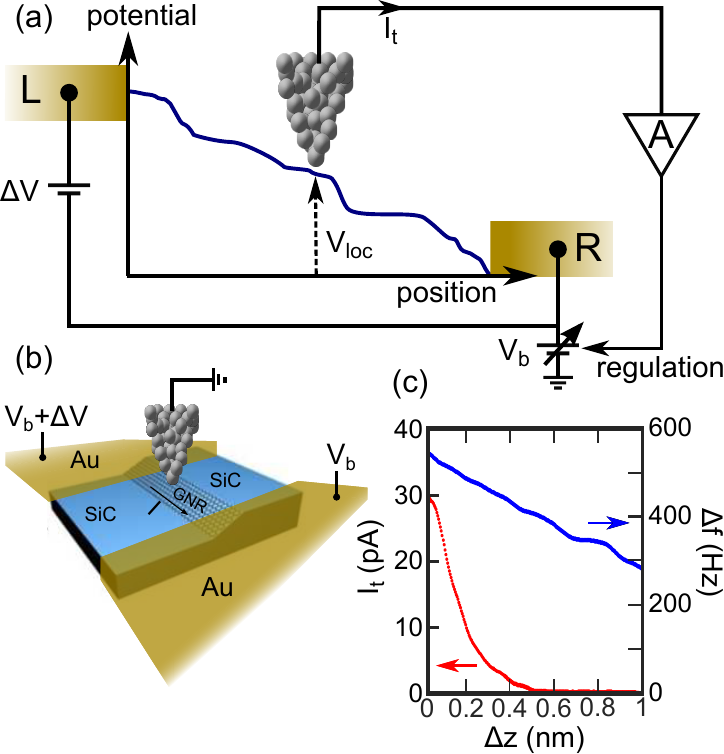}
\caption{(a) Operating principle of force-feedback scanning tunneling potentiometry. A voltage gradient $\Delta V$ from a floating voltage source is applied along a conducting sample confined between two metallic electrodes (L, R). A conducting tip (grounded via the tunneling current preamplifier) is scanned over the sample and the measurement of the local potential $V_{\rm loc}(x,y)$ is provided by the value of the additional regulation  voltage $V_b$ for which $I_{t}^{dc}=0$. The tip height regulation is performed simultaneously and independently by FM-AFM regulation. (b) Schematics of the graphene nanoribbon potentiometry experimental setup. (c) Experimental dependence of frequency shift $\Delta f$ (blue) and tunnel current $I_{t}$ (red) on the tip-sample distance, measured on a gold thin film (tip oscillation amplitude $A=30$ pm, $\Delta V=15$ mV, $V_b=0$).}
\label{fig:Principle}
\end{figure}

A combined AFM-STM setup \cite{Senzier2007,Samaddar2016} was used to measure local resistances (STM probe brought in hard contact with  graphene) and local potentiometry (using tunnel contact). In the local resistance method we scan the nanoribbon surface in AFM mode and lower the tip at regular intervals  by several nanometers, until a saturation of the tunnel current $I_{\rm t}$ is reached, at a value $I_{\rm t}^{\rm sat}$ (see Supp. Info). In agreement with prior reports, the hard contact does not damage  the ribbon or the SiC substrate \cite{Baringhaus2014,Aprojanz2018}. The single tip used here is grounded through the tunnel current preamplifier  (\mbox{Fig. \ref{fig:Principle}a}). Both gold contacts are brought at {\em the same} potential $V_b=10$ mV and serve as a counter-electrode. The two-point local resistance between the tip position and the counter-electrodes is defined as $R_{\rm loc}=V_b / I_t^{\rm sat}$. Note that $R_{\rm loc}$ results from two parallel resistances, extending from the tip-nanoribbon contact to either gold contact. 

For the local potential measurements, we developed a new scanning potentiometry method that we name force-feedback scanning tunneling potentiometry (FF-STP). It is adapted to the present case of a small conductive wire (or device) on an insulating substrate. High-resolution studies of the surface electrochemical potential are usually performed with scanning tunneling potentiometry. The latter is based on STM instrumentation and the feedback for stabilizing the tip height is provided by the tunnel current or conductance, which restricts measurements to fully conductive surfaces.  For insulating substrates, AFM-based potentiometry measurements are frequently used, such as conductive-AFM or electrostatic force microscopy (EFM) and derived techniques. However, these AFM-based techniques probe the {\em electrostatic} rather than the {\em electrochemical} potential, and are further far from providing comparable potential and spatial resolution \cite{Woodside2002} as STM-based techniques \cite{Willke2015,Wang2013,Xie,Yoshimoto2007,Bannani2008}.

For potentiometry experiments, the combined AFM-STM setup is operated in the non-contact AFM mode using a length extension resonator (LER) \cite{Samaddar2016, Samaddar2016b}. When operating at very small tip heights, which manifests as a large shift $\Delta f>100$ Hz of the AFM resonator frequency, a tunnel current can flow between the metallic tip and the conductive sample, as seen in Fig. \ref{fig:Principle}c. A fixed potential difference $\Delta V=15$ mV is applied along the sample from a battery, resulting in a position-dependent local potential $|V_{\rm loc}(x,y)|<|\Delta V|$ with respect to the lower potential electrode. To this, an overall sample bias $V_b$ is added, thereby shifting the sample potential as a whole. At a fixed tip position and height, controlled by the AFM signal, a second feedback loop adjusts $V_{\rm b}$ in order to maintain the \mbox{d.c.} tunnel current $I_{\rm t}=(V_{\rm loc}+V_{\rm b})/R_{\rm t}$ at zero ($R_{\rm t}$ is the tunneling resistance). Note that the a.c. part of the current, due to the oscillating tip height around its mean value at a frequency of 1 MHz, is averaged out by the current amplifier. At equilibrium, the local potential $V_{\rm loc}(x,y)$ is therefore the opposite of the feedback voltage $V_{\rm b}$. This Wheatstone-bridge-type potentiometry method does not require any assumption on the sample or tip local density of states, neither on the tip-sample tunnel resistance, as long as is not too large (in which case almost no tunneling occurs and the feedback voltage $V_b$ is ill-defined, see Supp. Info). Although the $V_b$-regulation becomes unstable on the non-conductive regions of the sample, the tip-to-sample distance remains the same, as it is independently controlled by the AFM signal. More on the technique can be found in Supp. Info.

\begin{figure}[t]
\includegraphics[width=0.65\columnwidth]{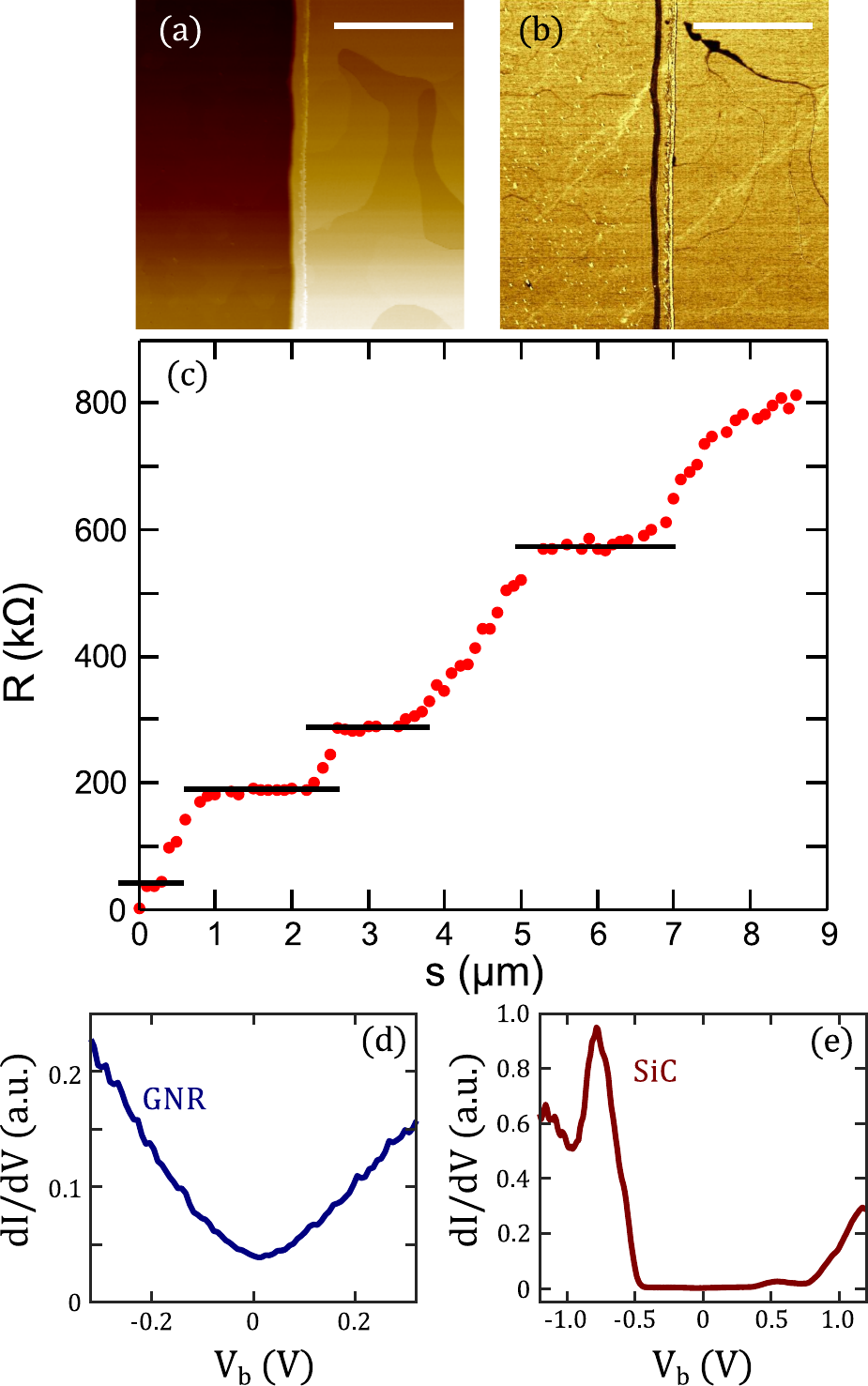}
\caption{
(a) AFM topography image showing the  sidewall mesa step and some natural curved SiC steps (height about 1 nm) (scale bar: 1 $\mu$m). (b) Lateral Force Microscopy scan of the same area showing a different friction contrast for graphene grown on the sidewall (black) compared to the terraces. 
(c) Local contact to ribbon resistance as a function of the curvilinear abscissa $s$; the gold contact is at  $s=0$.  Several larger that $1\,\mu$m-long segments show nearly constant local resistances, with $dR_{\rm loc}/ds \leq 10$ k$\Omega / \mu$m.
(d-e)  Low-temperature tunneling spectra on (d) sidewall, displaying a neutral graphene-like density of states and (e) SiC (or buffer-layer) terraces, displaying a semiconducting density of state ($T = 4$ K).}
\label{fig:plateau}
\end{figure}

We now move to measurements of the local resistance of individual sidewall ribbons, grown on the mesa shown in \mbox{Fig. \ref{fig:plateau}}a\&b. Similarly to Ref. \cite{Baringhaus2014}, scanning tunneling spectroscopy (STS) presented in \mbox{Fig. \ref{fig:plateau}}d  shows that the ribbon is essentially charge neutral, with a single sharp $dI/dV$ minimum at the Fermi level $\pm 20$ meV. This contrasts with STS on the terraces (Fig. \ref{fig:plateau}e), where the expected semiconducting SiC density of states is instead observed, with a marked gap spanning -500 to +700 meV. 
\mbox{Fig. \ref{fig:plateau}}c shows the local resistance 
as a function of the curvilinear abscissa $s$ starting from one contact. Several features stand out: a resistance jump of about $40$ k$\Omega$, from the first point of measurement on the gold contact to the ribbon, as well as four well defined flat plateaus over several $\mu$m-long segments and transition regions between them. The presence of resistance plateaus and the value of the contact resistance compare well with what was
 previously observed in similar ribbons at charge neutrality \cite{Baringhaus2014}. 
 
Following the analysis in Ref. \cite{Baringhaus2014}, the ribbon resistance at charge neutrality can be written as $R(s)=\alpha(h/e^2)(1+s/\lambda_0)$, where $s$ is the distance from the contact and $\lambda_0$ the mean free path.  The 40 k$\Omega$ contact resistance can be interpreted as a single conductance channel \cite{Baringhaus2014} with a transmission $\alpha=0.65$. The variation of $R_{\rm loc}$ with distance $s$ on the flat plateaus is below experimental sensitivity.  Given the experimental noise, an upper bound of $dR_{\rm loc}/ds$ on the plateaus is $\sim 10$ k$\Omega / \mu$m, which corresponds to $\lambda_0>3.3\,\mu$m. These room temperature mean path values are remarkably large, all the more given the period of time between ribbon production and the measurement (several months), the exposure to the environment, the contact deposition process, the absence of post-processing high-temperature treatment and measurement in non-UHV conditions. These results  confirm and expand those found in the extremely well controlled experimental conditions (in UHV or on alumina protected ribbons) \cite{Baringhaus2014} and demonstrate the robustness of the ballistic transport in these systems.

In between plateaus, steep $dR_{\rm loc}/ds$ slopes are found, reminiscent of the exponential increase of $R_{\rm loc}(s)$ at large distances, reported in Ref.\cite{Baringhaus2014}. Electronic edge state scattering on single-point defects is expected to produce local resistance jumps of the order of  $h/e^2$ at the location of the scattering center \cite{Baringhaus2014, Datta95}, which is observed in several instances (see \mbox{Fig. \ref{fig:plateau}}c). Elsewhere, seemingly continuous slopes are observed. These might be related to a change in the material itself and the destruction of the conducting edge state at these locations, leading to the large resistivity observed \cite{Baringhaus2015}. Alternatively, they could be following the same pattern as the onset of exponential increase of clean ribbons, but on shorter distances. The resurgence of several plateaus is clearly special and worth further investigation. 

\begin{figure}[!t]
\vspace{-1cm}
\includegraphics[width=0.7\columnwidth]{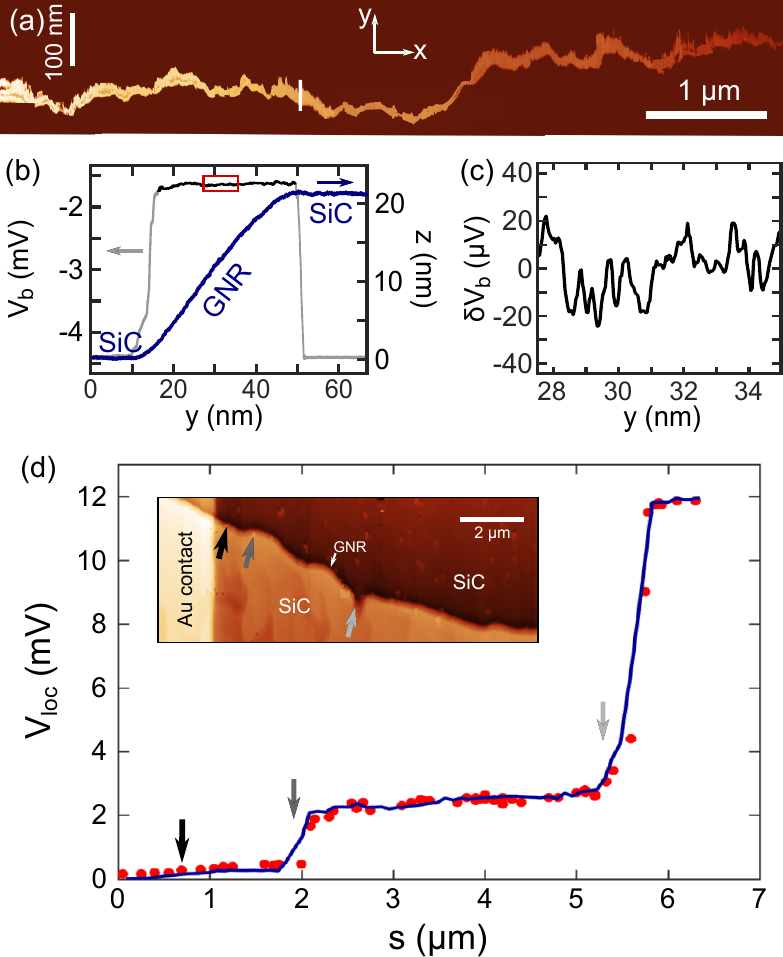}
\caption{(a) Local potential map of a voltage-biased 40 nm-wide and $\sim30$ $\mu$m-long graphene nanoribbon. The overall voltage drop along the nanoribbon segment presented here is about 4 mV. (b) Simultaneously acquired potential and topography profiles in the transverse direction of the nanoribbon. The conductive nanoribbon is on the inclined facet, confined between two insulating SiC terraces. Outside of the inclined facet (grey) the $V_b$ regulation is saturated. (c) Zoomed-in view of the $V_{b}$ signal, highlighting variations around its local average value. (d) Red dots: local potential profile measured by FF-STP of another ribbon, averaged over the transverse direction ($\Delta V=15$ mV and $\Delta f=300$ Hz). The blue line shows the  local potential profile calculated from local resistance  measurement of the same segment in hard-contact mode, with $V_{b}=10$ mV (see text). (inset) AFM topography of the measured graphene nanoribbon in the vicinity of a gold contact. Three prominent defects are highlighted by wide arrows. Feedback parameters (with $p$ the respective proportional gains and $\tau$ the integration times): $V_b$ feedback: $p_V = 8$ V/$\mu$A, $\tau_V = 100$ $\mu$s; frequency modulation AFM: $A$ = 15 pm, $\Delta f$ = 600 Hz, $p_{\rm AFM}=1$ V/$\mu$m, $\tau_{\rm AFM} = 300$ $\mu$s ; height feedback: $p_z = 2$ pm/Hz, $\tau_z= 1$ ms. The scan speed is 1 nm/s.}

\label{ribbon}
\end{figure}

The local resistance was observed with hard contacts; we now turn to the non-invasive potential measurements. 
Figure \ref{ribbon}a shows an example of high-resolution local electrochemical potential map of a $\sim 10\, \mu$m-long section of a meandering nanoribbon obtained with a fixed voltage gradient $\Delta V=15$ mV  set between the two gold contacts by a floating voltage source. For better readability, the $x$-axis scale is strongly compressed compared to the $y$ axis and the nanoribbon is in reality less meandering than it seems from the map. Even in the case of a very irregular conducting trace, the transverse line cut of (\mbox{Fig. \ref{ribbon}b})  shows that the conducting region  is confined to the inclined sidewall facet between two SiC(0001) terraces. Outside of the nanoribbon (grey part of the $V_{\rm b}$ curve), the regulation saturates to a physically irrelevant  value a few mV below the local ribbon potential value (see discussion in Supp. Info.). Nevertheless, as soon as the tip returns to the conducting regions, the regulation is immediately operative and provides an accurate measure of the local potential $V_{\rm loc}=-V_{\rm b}$. Note that within the experimental noise level $\sim 10 \, \mu$V, the potential is position-independent in the transverse direction to the ribbon, see \mbox{Fig. \ref{ribbon}c}.

 The local potential $V_{\rm loc}$ gradually drops  along the nanoribbon main axis. However the potential drop is not monotonous, and extended plateau regions are observed in most measured ribbons where a comparatively flat potential landscape ($dV_{\rm loc}/ds < 100$ $\mu$V/$\mu$m) is measured over $\mu$m-long distances. \mbox{Fig. \ref{ribbon}}d shows an example of a $V_{\rm loc}(s)$ trace where long plateaus are observed, similarly to \mbox{Fig. \ref{fig:plateau}}c,  despite an irregular topographic structure and the presence of contaminants (see \mbox{Figs. \ref{ribbon}}d, inset).

 Local resistance measurements were also performed on the same nanoribbon segment where the scanning potentiometry was measured. The two experimental quantities can be related via elementary circuit equations,
\begin{equation}
V_{\rm calc}=\frac{\Delta V}{2} \left( 1\pm \sqrt{1-4\,\frac{R_{\rm loc}}{R_{\rm tot}}   } \right).
\end{equation}

The calculated local potential can be adjusted to the measured one with a single free parameter that is the  total resistance of the nanoribbon $R_{\rm tot}$. An excellent agreement is found between  $V_{\rm loc}(s)$ and $V_{\rm calc}(s)$ (black trace in \mbox{Fig. \ref{ribbon}}d) with $R_{\rm tot}=72$ M$\Omega$ (see Supp. Info). It is quite remarkable that despite the overall rather high ribbon resistance, $\mu$m-long segments have a distance dependence of the local potential below experimental resolution.
The high ribbon resistance is most probably caused by isolated defects,
such as the kink near $s\approx 5.5 \, \mu$m (light grey arrow), ribbon irregularities and adsorbates.
 
Most importantly, the excellent correlation between the experimental and calculated traces indicates that the local potential as measured by our technique accurately reflects  the variations of the local resistance, but without the invasiveness of a hard metal tip pressed against the graphene.  Graphene is a textbook case as it is remarkably resistant to hard contact pressure that neither tears graphene nor leaves deposits on it. The equivalence of both potentiometry and resistance techniques for graphene indicates that the local potentiometry method may be used for much more delicate materials, where hard contact would be destructive.

We have looked for structural features at the boundaries of the $V_{\rm loc}(s)$ plateau regions that would signal a change of transport regime. In some cases, such as indicated by the light grey arrow in the inset of \mbox{Fig. \ref{ribbon}}d, a large topographic disruption of the mesa is observed where the $V_{\rm loc}(s)$  plateau terminates. In other instances, more localized defects (white dot) is seen (black and grey arrows). The $V_{\rm loc}(s)$ plateaus are observed for clean edges (no white dots), yet they appear immune to the sidewall orientation (see for instance between the light and dark grey arrows). Even very irregular ribbons such as the one shown in \mbox{Fig. \ref{ribbon}}a show longer than $\mu$m-long plateaus (6 $\mu$m-long plateaus are shown for that ribbon - see Supp. Info). 

In conclusion, the quasi distance-independent potential and resistance plateaus observed in epigraphene sidewall nanoribbons confirm larger than $\mu$m long mean free paths, even though the ribbons were processed with contacts and not measured in UHV. The robustness of the ballistic transport under these conditions, and for meandering ribbons, is remarkable. This is an important step towards large scale use of these ribbons for electronic applications. The new high-resolution potentiometry and local resistance measurements developed here provide consistent information, despite their quite different nature and degree of invasiveness. Applicable to a mixed conducting / insulating surface, the force-feedback technique expands the STP approach to the field of nano-electronic devices and, considering its non invasive character, it can be used to probe high resolution electrical characteristic of a large variety of materials.    

\medskip
We acknowledge funding from the from the European Union under the Marie Sklodowska-Curie Grant Agreement 766025, the US AFOSR (FA9550-13-0217), the NSF-ECCS (1506006) and NSF-DMR (1308835). C. B. and V. P. acknowledge funding from the European Union grant agreements 696656 and 785219. This work was also made possible by the French-American Cultural Exchange council through a Partner University Fund and a Thomas Jefferson grant. Dogukan Deniz is warmly thanked for making the sidewall ribbon sample. The devices were processed at the Nanofab platform at Institut N\'eel with the help of B. Fernandez.

\newpage

\section{\bf Supplemental Information File: Non-invasive nanoscale potentiometry and ballistic transport in epigraphene nanoribbons} 

\setcounter{figure}{0}
\makeatletter
\renewcommand{\theequation}{S\arabic{equation}}
\renewcommand{\thefigure}{S\arabic{figure}}
\global\long\def\theequation{S\arabic{equation}}
\global\long\def\thefigure{S\arabic{figure}}
\renewcommand{\thetable}{S\arabic{table}}

\bigskip

	This supporting information file discusses details of the Force-feedback scanning tunneling potentiometry, sample preparation, extended plateaus on an irregular ribbon, sample characterisation, and ribbon resistance measurements.

\bigskip

{\it Force-feedback scanning tunneling potentiometry} - In our setup, the tip height is controlled in the frequency-modulation mode, using a length extension resonator at a frequency $f_{\rm res}\approx1$ MHz \cite{Heike2003,An2005,Giessibl2011}. The tip-sample distance varies periodically at the resonance frequency with an amplitude $A$, but the resulting time-dependent current is beyond the bandwidth of the current amplifier. The tip, made of a sharp 4 $\mu$m-diameter W wire shaped by focused ion beam \cite{Samaddar2016}, is approached to the surface by increasing the set point frequency shift $\Delta f$. When reaching values of several 100 Hz, a tunnel resistance in the 1 G$\Omega$ range is established. In order to quantify the potentiometric resolution, we have performed measurements on an evaporated Au thin film. In the absence of externally imposed potential gradients, the \mbox{r.m.s.} fluctuations $\delta V_{\rm loc}$ are found in the 10 $\mu$V range for an acquisition time of 40 ms, which is equivalent to a noise level of about $2\,\mu$V/$\sqrt{\rm Hz}$, only slightly larger than the associated Johnson noise of the tunnel junction.

Due to the non-linear ribbon-tip distance dependence of the tunnel current, its time average can differ from its value for static conditions. Assuming an exponential decay of the tunnel current with decay length $\tilde{z}$, this can be translated into a renormalised tunnel resistance, with respect to its static value, following $R_{\rm t}(A)=R_{\rm t}(0)/J_{\rm 0}(iA/\tilde{z})$, where $J_{\rm 0}$ is the zeroth-order Bessel function and $A$ the tip oscillation amplitude. We keep $A$ smaller than the decay length of the tunnel current $\tilde{z}\approx 100$ pm. Therefore, the tunneling resistance renormalization effect is small. Furthermore, as stressed above, the determination of $V_{\rm loc}$ is actually insensitive to the precise value of $R_{\rm t}$.

For the local potentiometry to be non-invasive, $R_t$ must in principle be larger than the internal sample resistance. Yet, if the effects of the \mbox{a.c.} current can be neglected, the fact that the \mbox{d.c.} current to the tip is zero strongly relaxes this constraint. 

As to limit the divergence of $V_{\rm b}$ in the non-conductive regions, we beforehand set an admissible voltage range on the order of $\Delta V$, centered on the physically relevant potential window. 

\bigskip

{\it Sample preparation} - Epigraphene samples were produced by the confinement-controlled sublimation method by thermal decomposition of 4H-SiC \cite{deHeer2011}. Trenches  25 nm deep were produced in into SiC(0001) by lithography patterning and reactive ion  plasma etching. Annealing at 1250$\ensuremath{^\circ}$ C for 20 min followed by 1420$\ensuremath{^\circ}$ C for 10 minutes resulted in a set of parallel facets (\={1} \={1} 2 15)  orientation onto which graphene preferentially grows, producing an array of parallel 1 mm long graphene nanoribbons of width about 40 to 65 nm.

\bigskip

\begin{figure}[!t]
\includegraphics[width=0.5\columnwidth]{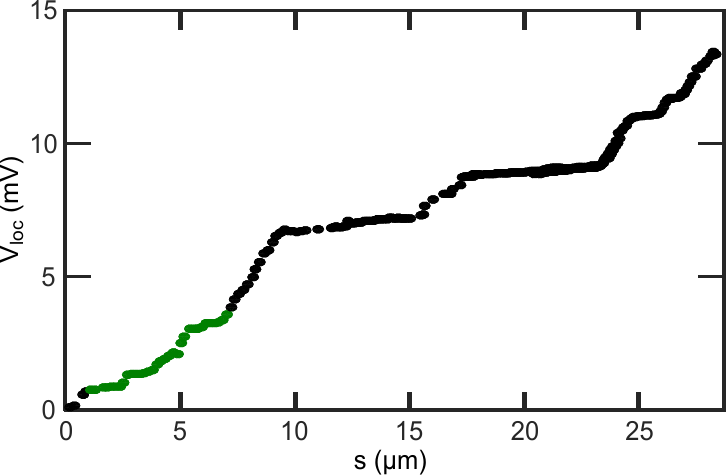}
\caption{Potential profile $V_{\rm loc}$ (averaged over the transverse direction)  as a function of the curvilinear abscissa $s$ along the entire ribbon, shown in Fig. 3 (main manuscript). The high-resolution map shown in Fig. 3(a) corresponds to the area of green dots.}
\label{ribbon-supp}
\end{figure}

{\it Extended plateaus on an irregular ribbon} - Very long $V(s)$ plateaus are observed even for quite irregular ribbons such as the one presented in Fig. 3 (main text). An example is given in Fig. \ref{ribbon-supp}, where flat potential are observed on more that 7 $\mu$m-long segments.

\bigskip

{\it Sample characterization} - After growth, the graphene nanoribbons are inspected by room-temperature AFM in order to investigate the overall shape of the nanoribbons and the flatness of the SiC terraces (\mbox{Fig.} 2a in main manuscript). Room-temperature Lateral Force Microscopy measurements show a distinctive contrast for the sidewall region (\mbox{Fig.} 2b), where the lower friction identifies graphene.

\bigskip

{\it Ribbon resistance} - For  all the ribbons, the two-probe $I(V)$ characteristics measured in a room-temperature probe station are linear up to a bias voltage of $\pm$ 10 V,  indicating  good quality of the metal/graphene interface and the absence of prominent Schottky barriers. 

The total  2-point resistance of the samples is  of the order of 100 k$\Omega$ to 1 M$\Omega$. Since about 100 such ribbons are connected in parallel, the resistance of individual ribbons cannot be determined from transport measurements alone. Assuming all the ribbons are equivalent would give individual ribbon resistance of 10 -- 100 M$\Omega$, consistent with the value of 72 M$\Omega$ from the potential fit in  \mbox{Fig. 3}. Although large, this resistance is temperature independent down to 4 K. These values are much larger that the 25 -- 100 k$\Omega$ of UVH-annealed ribbons, related to the steep $R(s)$ slopes in-between the plateaus. 
Note that the single channel ballistic ribbons in ref.\cite{Baringhaus2014} have 26 k$\Omega$ resistances that increase sharply after about 15 $\mu$m, so that the sharp rise is not unexpected, but occurs after shorter plateaus.

\begin{figure}[t]
\includegraphics[width=0.95\columnwidth]{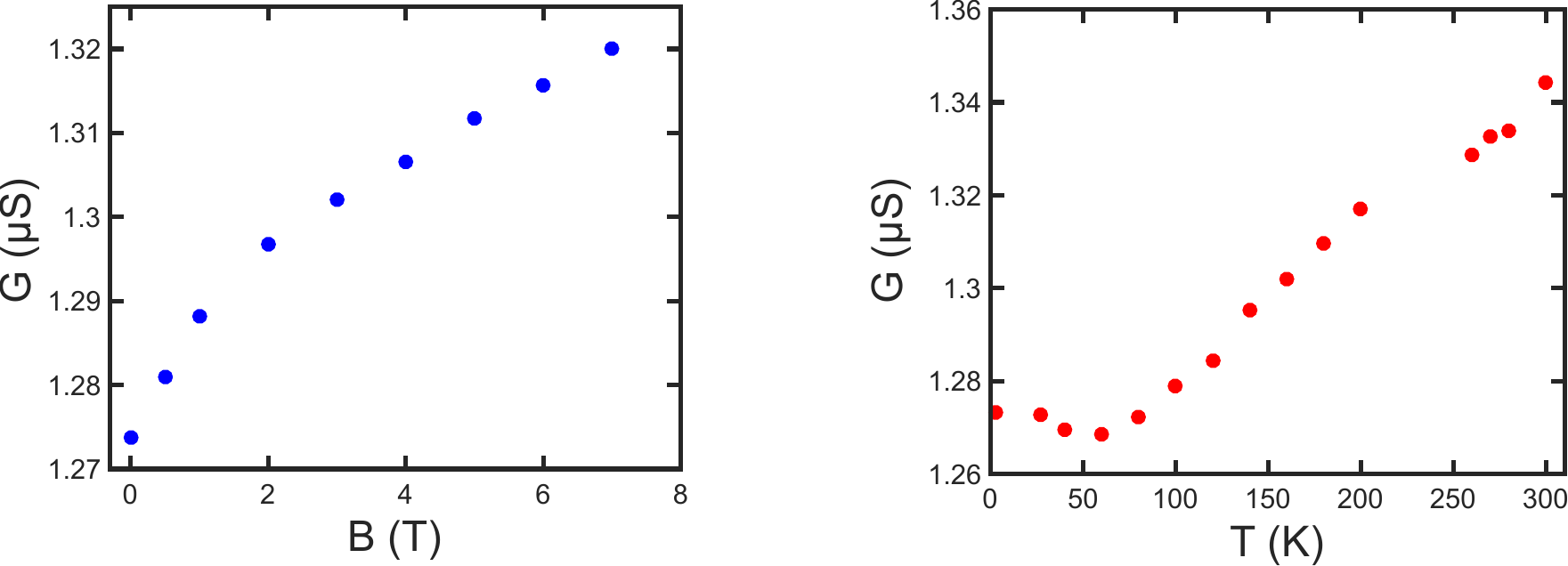}
\caption{(a) Measured 2-point conductance as a function of temperature (T = 2.5 to 300 K, B = 0), showing only a slight decrease as temperature is lowered, with a slight saturation below 50 K. (b) Measured 2-point conductance as a function of transverse magnetic field (B = 0 to 7 T, T = 2.5 K).}
\label{fig:conductance}
\end{figure}

Magneto-transport were measured in a dedicated setup reaching 2.5 K and 7 T for selected nanoribbons. The measured 2-points conductance is quasi temperature independent, with only a $\sim$5 \% decrease from 300 K to 4 K (Fig. S2), demonstrating that graphene nanoribbons display a non-semiconducting behaviour. This is consistent with previous experiments of other sidewall curved graphene nanoribbons \cite{Baringhaus2014,Ruan2012} but in sharp contrast with patterned ribbons from exfoliated graphene \cite{Oostinga2010}. The 2 point-resistance also slightly decreases for increasing perpendicular magnetic field (\mbox{Fig.} \ref{fig:conductance}b), in agreement with the magnetoresistance at low magnetic field in other graphene sidewall nanoribbons \cite{Baringhaus2014}.

\bibliography{Bibliography}

\providecommand{\latin}[1]{#1}
\providecommand*\mcitethebibliography{\thebibliography}
\csname @ifundefined\endcsname{endmcitethebibliography}
  {\let\endmcitethebibliography\endthebibliography}{}
\begin{mcitethebibliography}{30}
\providecommand*\natexlab[1]{#1}
\providecommand*\mciteSetBstSublistMode[1]{}
\providecommand*\mciteSetBstMaxWidthForm[2]{}
\providecommand*\mciteBstWouldAddEndPuncttrue
  {\def\EndOfBibitem{\unskip.}}
\providecommand*\mciteBstWouldAddEndPunctfalse
  {\let\EndOfBibitem\relax}
\providecommand*\mciteSetBstMidEndSepPunct[3]{}
\providecommand*\mciteSetBstSublistLabelBeginEnd[3]{}
\providecommand*\EndOfBibitem{}
\mciteSetBstSublistMode{f}
\mciteSetBstMaxWidthForm{subitem}{(\alph{mcitesubitemcount})}
\mciteSetBstSublistLabelBeginEnd
  {\mcitemaxwidthsubitemform\space}
  {\relax}
  {\relax}

\bibitem[Sprinkle \latin{et~al.}(2010)Sprinkle, Ruan, Hu, Hankinson, Rubio-Roy,
  Zhang, Wu, Berger, and De~Heer]{Sprinkle2010}
Sprinkle,~M.; Ruan,~M.; Hu,~Y.; Hankinson,~J.; Rubio-Roy,~M.; Zhang,~B.;
  Wu,~X.; Berger,~C.; De~Heer,~W.~A. \emph{Nature Nanotechnol.} \textbf{2010},
  \emph{5}, 727--731\relax
\mciteBstWouldAddEndPuncttrue
\mciteSetBstMidEndSepPunct{\mcitedefaultmidpunct}
{\mcitedefaultendpunct}{\mcitedefaultseppunct}\relax
\EndOfBibitem
\bibitem[Ruan \latin{et~al.}(2012)Ruan, Hu, Guo, Dong, Palmer, Hankinson,
  Berger, and De~Heer]{Ruan2012}
Ruan,~M.; Hu,~Y.; Guo,~Z.; Dong,~R.; Palmer,~J.; Hankinson,~J.; Berger,~C.;
  De~Heer,~W.~A. \emph{MRS Bulletin} \textbf{2012}, \emph{37}, 1138--1147\relax
\mciteBstWouldAddEndPuncttrue
\mciteSetBstMidEndSepPunct{\mcitedefaultmidpunct}
{\mcitedefaultendpunct}{\mcitedefaultseppunct}\relax
\EndOfBibitem
\bibitem[Baringhaus \latin{et~al.}(2014)Baringhaus, Ruan, Edler, Tejeda, Sicot,
  Taleb-Ibrahimi, Li, Jiang, Conrad, Berger, Tegenkamp, and
  de~Heer]{Baringhaus2014}
Baringhaus,~J.; Ruan,~M.; Edler,~F.; Tejeda,~A.; Sicot,~M.; Taleb-Ibrahimi,~A.;
  Li,~A.-P.; Jiang,~Z.; Conrad,~E.~H.; Berger,~C.; Tegenkamp,~C.;
  de~Heer,~W.~A. \emph{Nature} \textbf{2014}, \emph{506}, 349--354\relax
\mciteBstWouldAddEndPuncttrue
\mciteSetBstMidEndSepPunct{\mcitedefaultmidpunct}
{\mcitedefaultendpunct}{\mcitedefaultseppunct}\relax
\EndOfBibitem
\bibitem[Aprojanz \latin{et~al.}(2018)Aprojanz, Power, Bampoulis, Roche, Jauho,
  Zandvliet, Zakharov, and Tegenkamp]{Aprojanz2018}
Aprojanz,~J.; Power,~S.~R.; Bampoulis,~P.; Roche,~S.; Jauho,~A.-P.;
  Zandvliet,~H.~J.; Zakharov,~A.~A.; Tegenkamp,~C. \emph{Nature Commun.}
  \textbf{2018}, \emph{9}, 1--6\relax
\mciteBstWouldAddEndPuncttrue
\mciteSetBstMidEndSepPunct{\mcitedefaultmidpunct}
{\mcitedefaultendpunct}{\mcitedefaultseppunct}\relax
\EndOfBibitem
\bibitem[Baringhaus \latin{et~al.}(2015)Baringhaus, Aprojanz, Wiegand, Laube,
  Halbauer, H{\"u}bner, Oestreich, and Tegenkamp]{Baringhaus2015}
Baringhaus,~J.; Aprojanz,~J.; Wiegand,~J.; Laube,~D.; Halbauer,~M.;
  H{\"u}bner,~J.; Oestreich,~M.; Tegenkamp,~C. \emph{Appl. Phys. Lett.}
  \textbf{2015}, \emph{106}, 043109\relax
\mciteBstWouldAddEndPuncttrue
\mciteSetBstMidEndSepPunct{\mcitedefaultmidpunct}
{\mcitedefaultendpunct}{\mcitedefaultseppunct}\relax
\EndOfBibitem
\bibitem[Han \latin{et~al.}(2010)Han, Brant, and Kim]{Han2010}
Han,~M.~Y.; Brant,~J.~C.; Kim,~P. \emph{Phys. Rev. Lett.} \textbf{2010},
  \emph{104}, 056801\relax
\mciteBstWouldAddEndPuncttrue
\mciteSetBstMidEndSepPunct{\mcitedefaultmidpunct}
{\mcitedefaultendpunct}{\mcitedefaultseppunct}\relax
\EndOfBibitem
\bibitem[Gallagher \latin{et~al.}(2010)Gallagher, Todd, and
  Goldhaber-Gordon]{Gallagher2010}
Gallagher,~P.; Todd,~K.; Goldhaber-Gordon,~D. \emph{Phys. Rev. B}
  \textbf{2010}, \emph{81}, 115409\relax
\mciteBstWouldAddEndPuncttrue
\mciteSetBstMidEndSepPunct{\mcitedefaultmidpunct}
{\mcitedefaultendpunct}{\mcitedefaultseppunct}\relax
\EndOfBibitem
\bibitem[Dr{\"o}scher \latin{et~al.}(2011)Dr{\"o}scher, Knowles, Meir, Ensslin,
  and Ihn]{Droscher2011}
Dr{\"o}scher,~S.; Knowles,~H.; Meir,~Y.; Ensslin,~K.; Ihn,~T. \emph{Phys. Rev.
  B} \textbf{2011}, \emph{84}, 073405\relax
\mciteBstWouldAddEndPuncttrue
\mciteSetBstMidEndSepPunct{\mcitedefaultmidpunct}
{\mcitedefaultendpunct}{\mcitedefaultseppunct}\relax
\EndOfBibitem
\bibitem[Frank \latin{et~al.}(1998)Frank, Poncharal, Wang, and
  De~Heer]{Frank1998}
Frank,~S.; Poncharal,~P.; Wang,~Z.; De~Heer,~W.~A. \emph{Science}
  \textbf{1998}, \emph{280}, 1744--1746\relax
\mciteBstWouldAddEndPuncttrue
\mciteSetBstMidEndSepPunct{\mcitedefaultmidpunct}
{\mcitedefaultendpunct}{\mcitedefaultseppunct}\relax
\EndOfBibitem
\bibitem[Li \latin{et~al.}(2016)Li, Niquet, and Delerue]{Li2016}
Li,~J.; Niquet,~Y.-M.; Delerue,~C. \emph{Phys. Rev. Lett.} \textbf{2016},
  \emph{116}, 236602\relax
\mciteBstWouldAddEndPuncttrue
\mciteSetBstMidEndSepPunct{\mcitedefaultmidpunct}
{\mcitedefaultendpunct}{\mcitedefaultseppunct}\relax
\EndOfBibitem
\bibitem[Crossno \latin{et~al.}(2016)Crossno, Shi, Wang, Liu, Harzheim, Lucas,
  Sachdev, Kim, Taniguchi, Watanabe, Ohki, and Chung~Fong]{Crossno2016}
Crossno,~J.; Shi,~J.~K.; Wang,~K.; Liu,~X.; Harzheim,~A.; Lucas,~A.;
  Sachdev,~S.; Kim,~P.; Taniguchi,~T.; Watanabe,~K.; Ohki,~T.~A.;
  Chung~Fong,~K. \emph{Science} \textbf{2016}, \emph{351}, 1058--1061\relax
\mciteBstWouldAddEndPuncttrue
\mciteSetBstMidEndSepPunct{\mcitedefaultmidpunct}
{\mcitedefaultendpunct}{\mcitedefaultseppunct}\relax
\EndOfBibitem
\bibitem[Bandurin \latin{et~al.}(2016)Bandurin, Torre, Kumar, Shalom, Tomadin,
  Principi, Auton, Khestanova, Novoselov, Grigorieva, Ponomarenko, Geim, and
  Polini]{Bandurin2016}
Bandurin,~D.; Torre,~I.; Kumar,~R.~K.; Shalom,~M.~B.; Tomadin,~A.;
  Principi,~A.; Auton,~G.; Khestanova,~E.; Novoselov,~K.; Grigorieva,~I.;
  Ponomarenko,~L.~A.; Geim,~A.~K.; Polini,~M. \emph{Science} \textbf{2016},
  \emph{351}, 1055--1058\relax
\mciteBstWouldAddEndPuncttrue
\mciteSetBstMidEndSepPunct{\mcitedefaultmidpunct}
{\mcitedefaultendpunct}{\mcitedefaultseppunct}\relax
\EndOfBibitem
\bibitem[Palacio \latin{et~al.}(2015)Palacio, Celis, Nair, Gloter, Zobelli,
  Sicot, Malterre, Nevius, de~Heer, Berger, Conrad, Taleb-Ibrahimi, , and
  Tejeda]{Palacio2015}
Palacio,~I.; Celis,~A.; Nair,~M.~N.; Gloter,~A.; Zobelli,~A.; Sicot,~M.;
  Malterre,~D.; Nevius,~M.~S.; de~Heer,~W.~A.; Berger,~C.; Conrad,~E.~H.;
  Taleb-Ibrahimi,~A.; ; Tejeda,~A. \emph{Nano Lett.} \textbf{2015}, \emph{15},
  182--189\relax
\mciteBstWouldAddEndPuncttrue
\mciteSetBstMidEndSepPunct{\mcitedefaultmidpunct}
{\mcitedefaultendpunct}{\mcitedefaultseppunct}\relax
\EndOfBibitem
\bibitem[Celis \latin{et~al.}(2016)Celis, Nair, Taleb-Ibrahimi, Conrad, Berger,
  De~Heer, and Tejeda]{Celis2016}
Celis,~A.; Nair,~M.; Taleb-Ibrahimi,~A.; Conrad,~E.; Berger,~C.; De~Heer,~W.;
  Tejeda,~A. \emph{Jour. of Phys. D} \textbf{2016}, \emph{49}, 143001\relax
\mciteBstWouldAddEndPuncttrue
\mciteSetBstMidEndSepPunct{\mcitedefaultmidpunct}
{\mcitedefaultendpunct}{\mcitedefaultseppunct}\relax
\EndOfBibitem
\bibitem[Senzier \latin{et~al.}(2007)Senzier, Luo, and Courtois]{Senzier2007}
Senzier,~J.; Luo,~P.~S.; Courtois,~H. \emph{Appl. Phys. Lett.} \textbf{2007},
  \emph{90}, 043114\relax
\mciteBstWouldAddEndPuncttrue
\mciteSetBstMidEndSepPunct{\mcitedefaultmidpunct}
{\mcitedefaultendpunct}{\mcitedefaultseppunct}\relax
\EndOfBibitem
\bibitem[Samaddar \latin{et~al.}(2016)Samaddar, Yudhistira, Adam, Courtois, and
  Winkelmann]{Samaddar2016}
Samaddar,~S.; Yudhistira,~I.; Adam,~S.; Courtois,~H.; Winkelmann,~C.
  \emph{Phys. Rev. Lett.} \textbf{2016}, \emph{116}, 126804\relax
\mciteBstWouldAddEndPuncttrue
\mciteSetBstMidEndSepPunct{\mcitedefaultmidpunct}
{\mcitedefaultendpunct}{\mcitedefaultseppunct}\relax
\EndOfBibitem
\bibitem[Woodside and McEuen(2002)Woodside, and McEuen]{Woodside2002}
Woodside,~M.~T.; McEuen,~P.~L. \emph{Science} \textbf{2002}, \emph{296},
  1098--1101\relax
\mciteBstWouldAddEndPuncttrue
\mciteSetBstMidEndSepPunct{\mcitedefaultmidpunct}
{\mcitedefaultendpunct}{\mcitedefaultseppunct}\relax
\EndOfBibitem
\bibitem[Willke \latin{et~al.}(2015)Willke, Druga, Ulbrich, Schneider, and
  Wenderoth]{Willke2015}
Willke,~P.; Druga,~T.; Ulbrich,~R.~G.; Schneider,~M.~A.; Wenderoth,~M.
  \emph{Nature Commun.} \textbf{2015}, \emph{6}, 1--5\relax
\mciteBstWouldAddEndPuncttrue
\mciteSetBstMidEndSepPunct{\mcitedefaultmidpunct}
{\mcitedefaultendpunct}{\mcitedefaultseppunct}\relax
\EndOfBibitem
\bibitem[Wang \latin{et~al.}(2013)Wang, Munakata, Rozler, and
  Beasley]{Wang2013}
Wang,~W.; Munakata,~K.; Rozler,~M.; Beasley,~M.~R. \emph{Phys. Rev. Lett.}
  \textbf{2013}, \emph{110}, 236802\relax
\mciteBstWouldAddEndPuncttrue
\mciteSetBstMidEndSepPunct{\mcitedefaultmidpunct}
{\mcitedefaultendpunct}{\mcitedefaultseppunct}\relax
\EndOfBibitem
\bibitem[Xie \latin{et~al.}(2017)Xie, Dreyer, Bowen, Hinkel, Butera, Krafft,
  and Mayergoyz]{Xie}
Xie,~T.; Dreyer,~M.; Bowen,~D.; Hinkel,~D.; Butera,~R.; Krafft,~C.;
  Mayergoyz,~I. \emph{AIP Advances} \textbf{2017}, \emph{7}, 125205\relax
\mciteBstWouldAddEndPuncttrue
\mciteSetBstMidEndSepPunct{\mcitedefaultmidpunct}
{\mcitedefaultendpunct}{\mcitedefaultseppunct}\relax
\EndOfBibitem
\bibitem[Yoshimoto \latin{et~al.}(2007)Yoshimoto, Murata, Kubo, Tomita,
  Motoyoshi, Kimura, Okino, Hobara, Matsuda, Honda, Katayama, and
  Hasegawa]{Yoshimoto2007}
Yoshimoto,~S.; Murata,~Y.; Kubo,~K.; Tomita,~K.; Motoyoshi,~K.; Kimura,~T.;
  Okino,~H.; Hobara,~R.; Matsuda,~I.; Honda,~S.-i.; Katayama,~M.; Hasegawa,~S.
  \emph{Nano Lett.} \textbf{2007}, \emph{7}, 956--959\relax
\mciteBstWouldAddEndPuncttrue
\mciteSetBstMidEndSepPunct{\mcitedefaultmidpunct}
{\mcitedefaultendpunct}{\mcitedefaultseppunct}\relax
\EndOfBibitem
\bibitem[Bannani \latin{et~al.}(2008)Bannani, Bobisch, and
  M{\"o}ller]{Bannani2008}
Bannani,~A.; Bobisch,~C.; M{\"o}ller,~R. \emph{Rev. Sci. Instrum.}
  \textbf{2008}, \emph{79}, 083704\relax
\mciteBstWouldAddEndPuncttrue
\mciteSetBstMidEndSepPunct{\mcitedefaultmidpunct}
{\mcitedefaultendpunct}{\mcitedefaultseppunct}\relax
\EndOfBibitem
\bibitem[Samaddar \latin{et~al.}(2016)Samaddar, Coraux, Martin, Gr{\'e}vin,
  Courtois, and Winkelmann]{Samaddar2016b}
Samaddar,~S.; Coraux,~J.; Martin,~S.~C.; Gr{\'e}vin,~B.; Courtois,~H.;
  Winkelmann,~C.~B. \emph{Nanoscale} \textbf{2016}, \emph{8},
  15162--15166\relax
\mciteBstWouldAddEndPuncttrue
\mciteSetBstMidEndSepPunct{\mcitedefaultmidpunct}
{\mcitedefaultendpunct}{\mcitedefaultseppunct}\relax
\EndOfBibitem
\bibitem[Datta(1997)]{Datta95}
Datta,~S. \emph{Electronic transport in mesoscopic systems}; Cambridge
  University Press, 1997\relax
\mciteBstWouldAddEndPuncttrue
\mciteSetBstMidEndSepPunct{\mcitedefaultmidpunct}
{\mcitedefaultendpunct}{\mcitedefaultseppunct}\relax
\EndOfBibitem
\bibitem[Heike and Hashizume(2003)Heike, and Hashizume]{Heike2003}
Heike,~S.; Hashizume,~T. \emph{Appl. Phys. Lett.} \textbf{2003}, \emph{83},
  3620--3622\relax
\mciteBstWouldAddEndPuncttrue
\mciteSetBstMidEndSepPunct{\mcitedefaultmidpunct}
{\mcitedefaultendpunct}{\mcitedefaultseppunct}\relax
\EndOfBibitem
\bibitem[An \latin{et~al.}(2005)An, Eguchi, Akiyama, and Hasegawa]{An2005}
An,~T.; Eguchi,~T.; Akiyama,~K.; Hasegawa,~Y. \emph{Appl. Phys. Lett.}
  \textbf{2005}, \emph{87}, 133114\relax
\mciteBstWouldAddEndPuncttrue
\mciteSetBstMidEndSepPunct{\mcitedefaultmidpunct}
{\mcitedefaultendpunct}{\mcitedefaultseppunct}\relax
\EndOfBibitem
\bibitem[Giessibl \latin{et~al.}(2011)Giessibl, Pielmeier, Eguchi, An, and
  Hasegawa]{Giessibl2011}
Giessibl,~F.~J.; Pielmeier,~F.; Eguchi,~T.; An,~T.; Hasegawa,~Y. \emph{Phys.
  Rev. B,} \textbf{2011}, \emph{84}, 125409\relax
\mciteBstWouldAddEndPuncttrue
\mciteSetBstMidEndSepPunct{\mcitedefaultmidpunct}
{\mcitedefaultendpunct}{\mcitedefaultseppunct}\relax
\EndOfBibitem
\bibitem[De~Heer \latin{et~al.}(2011)De~Heer, Berger, Ruan, Sprinkle, Li, Hu,
  Zhang, Hankinson, and Conrad]{deHeer2011}
De~Heer,~W.~A.; Berger,~C.; Ruan,~M.; Sprinkle,~M.; Li,~X.; Hu,~Y.; Zhang,~B.;
  Hankinson,~J.; Conrad,~E. \emph{Proc. Nat. Acad. Sci.} \textbf{2011},
  \emph{108}, 16900--16905\relax
\mciteBstWouldAddEndPuncttrue
\mciteSetBstMidEndSepPunct{\mcitedefaultmidpunct}
{\mcitedefaultendpunct}{\mcitedefaultseppunct}\relax
\EndOfBibitem
\bibitem[Oostinga \latin{et~al.}(2010)Oostinga, Sac{\'e}p{\'e}, Craciun, and
  Morpurgo]{Oostinga2010}
Oostinga,~J.~B.; Sac{\'e}p{\'e},~B.; Craciun,~M.~F.; Morpurgo,~A.~F.
  \emph{Phys. Rev. B} \textbf{2010}, \emph{81}, 193408\relax
\mciteBstWouldAddEndPuncttrue
\mciteSetBstMidEndSepPunct{\mcitedefaultmidpunct}
{\mcitedefaultendpunct}{\mcitedefaultseppunct}\relax
\EndOfBibitem
\end{mcitethebibliography}


\end{document}